\documentclass[eqsecnum,showkeys,showpacs,nofootinbib,aps]{revtex4}
\renewcommand{\theequation}{\arabic{equation}}
\def\bea{\begin{eqnarray}}
\def\eea{\end{eqnarray}}
\def\bwide{\begin{widetext}}
\def\ewide{\end{widetext}}

\newcommand{\nn}{\nonumber}

\def\beq{\begin{equation}}
\def\eeq{\end{equation}}

\begin{document}
\title{Higher dimensional flat embedding of Taub-NUT-AdS spacetime}
\author{Soon-Tae Hong}
\affiliation{Department of Science Education and Research Institute for Basic Sciences, Ewha Womans
University, Seoul 120-750 Korea}
\date{\today}%
\begin{abstract}
We construct a global flat embedding structure of a Taub-NUT-AdS spacetime to yield a (6+5)-dimensional 
novel global embedding Minkowski spacetime. We also investigate Taub-NUT, Schwarzschild-AdS and Schwarzschild limits of 
the global embedding by exploiting parameter reduction scheme. In particular, we observe in the 
vanishing monopole strength limit of the Taub-NUT-AdS that the parameter reduction is not smoothly 
applicable to the Schwarzschild-AdS, due to the presence of imaginary roots of its 
lapse function associated with event horizon. Moreover, reductions from the Taub-NUT-AdS and Schwarzschild-AdS 
to Taub-NUT and Schwarzschild, respectively, are successfully performed. Finally, we construct the 
global embedding Minkowski spacetimes for the patches inside the event horizons of the Taub-NUT-AdS and its extended manifolds. 
\end{abstract}
 \maketitle

\section{Introduction}
\setcounter{equation}{0}
\renewcommand{\theequation}{\arabic{section}.\arabic{equation}}

Since a Taub-NUT spacetime was proposed~\cite{taub51,nut63}, tremendous progress has been made in
general relativity community. The Taub-NUT was exploited to treat on the manifold its associated novel 
topics such as D-brane, instanton bundle and hyper-Kahler quotients~\cite{witten09}, rescaling of 
conformal Yano-Killing tensors~\cite{jezierski07}, moduli spaces of instantons and mirror symmetry~\cite{cherkis09}, 
geodesics and Weierstrass elliptic functions~\cite{kunz10}, 
black ring with dipole magnetic charges~\cite{bena11}, Gowdy symmetry and Cauchy horizon~\cite{beyer12}, Dirac 
operator~\cite{moroianu11,jante14} and integrability in conformally coupled gravity~\cite{bardoux14}. Next, 
the Taub-NUT-AdS has coordinate singularity occurring at certain values of the radial coordinate 
corresponding to bifurcate Killing horizons. The Taub-NUT-AdS includes the gravitational analogue of 
Yang-Mills instanton~\cite{hawking77}, the Kaluza-Klein monopole and gravitational instanton~\cite{manton85,atiyah85}, 
M-theory in higher dimensional spaces~\cite{awad02}, gravitomagnetic monopole source effects~\cite{bini03} 
and O(N) vector model on squashed three-sphere~\cite{jezierski072}.

On the other hand, a global embedding Minkowski space (GEMS) was intensively investigated by several
authors~\cite{kasner21,fronsdal59,friedman65,goenner80,rosen65,narnhofer96,chamblin97,deser97,deser98,kim00,
hong00,banerjee10,hong002,hong01,hong04,hong10,hong12,hong132,gems08,gems14}.
A thermal Hawking effect~\cite{hawking75} on a curved manifold can be looked 
at~\cite{chen04,santos04,tian05,radu06,langlois06,russo08} as
an Unruh effect~\cite{unruh76} in a flat higher-dimensional spacetime such as the GEMS. 
Some physical aspects of the GEMS are to remove coordinate singularity in the four-metrics and to 
define the Hawking temperature~\cite{hawking75,bekenstein73} and an entropy as area of the Rindler horizon in 
this formalism. Specifically, the GEMS scheme was adapted to 
the Ba\~nados-Teitelboim-Zanelli~\cite{btz,carlip95,cangemi93}, the Schwarzschild~\cite{sch16} 
together with its anti-de Sitter (AdS), the Reissner-Nordstr\"om (RN)~\cite{rn18,peca18} 
and the RN-AdS~\cite{kim00} models. Very recently the GEMS structure of the Taub-NUT spacetime was investigated 
to yield a (6+5)-dimensional manifold~\cite{hong10}.

In this paper, we will further analyze the GEMS of a Taub-NUT-AdS spacetime, by introducing an ansatz 
for coordinate transformations to handle a cross term of time and angle coordinates in its four-metric. 
Since the Taub-NUT-AdS possesses mass, cosmological constant and monopole strength as parameters, we will 
choose vanishing limits of these parameters to investigate Taub-NUT, Schwarzschild-AdS and Schwarzschild cases of 
the global embedding by exploiting parameter reduction scheme. We will show that the Taub-NUT-AdS lapse function has 
imaginary roots and thus, in its vanishing monopole strength limit, the parameter reduction cannot be successfully applied to the Schwarzschild-AdS. However, we will make reduction from the Taub-NUT-AdS and the Schwarzschild-AdS to the Taub-NUT and the Schwarzschild, respectively. In Section II, we construct a GEMS of a Taub-NUT-AdS spacetime.
In Section III we investigate Taub-NUT, Schwarzschild-AdS and Schwarzschild limits of 
the global embedding. Section IV includes summaries and discussions.

\section{Taub-NUT-AdS GEMS}
\setcounter{equation}{0}
\renewcommand{\theequation}{\arabic{section}.\arabic{equation}}


We start with the Taub-NUT-AdS four-metric given by \beq
ds^{2}=-\frac{\Delta}{\Sigma}(dt+2n\cos\theta d\phi)^{2}
+\frac{\Sigma}{\Delta}dr^2+\Sigma(d\theta^{2}+\sin^{2}\theta
d\phi^{2})\label{taubmetric}\eeq where
\bea
\Delta&=&r^{2}-2Mr-n^{2}+l^{-2}(r^{4}+6n^{2}r^{2}-3n^{4}),\label{delta}\\
\Sigma&=&r^{2}+n^{2},\label{sigma}
\eea
with the parameters $M$, $n$ and $l$ being associated with the mass, 
gravitational magnetic monopole strength of the source, and the cosmological constant. Introducing 
$r_{\pm}$, which are roots of $\Delta$, we can rewrite (\ref{delta}) for $r\geq r_{+}$ as 
\beq 
\Delta=l^{-2}(r-r_{+})(r-r_{-})\left[r^{2}+(r_{+}+r_{-})r-c\right],\label{delta2} \eeq
where 
\beq
c=\frac{n^{2}(3n^{2}+l^{2})}{r_{+}r_{-}}.
\eeq

Next, we evaluate the $l^{2}$ in terms of $r_{\pm}$ as follows
$$
l^{2}=\frac{(r_{+}+r_{-})(r_{+}^{2}+r_{-}^{2}+6n^{2})}{2M-r_{+}-r_{-}}.
$$
Moreover, we find an identity
\beq
3(2M+r_{+}+r_{-})n^{4}+[(r_{+}^{2}+r_{-}^{2})(r_{+}+r_{-})
+12Mr_{+}r_{-}]n^{2}+r_{+}r_{-}[2M(r_{+}^{2}+r_{-}^{2}+r_{+}r_{-})-r_{+}r_{-}(r_{+}+r_{-})]=0.\nn
\eeq
Finally we find the surface gravity $\kappa_{H}$ of the Taub-NUT-AdS spacetime
\beq
\kappa_{H}^{~}=\frac{(r_{+}-r_{-})(2r_{+}^{2}+r_{+}r_{-}-c)}{2l^{2}(r_{+}^{2}+n^{2})}.
\label{kappah}
\eeq

In order to figure out constructing the GEMS, which will be shown below, we rewrite 
the four-metric in (\ref{taubmetric}) as 
\beq
ds^{2}=-N^{2}dt^{2}-4nN^{2}\cos\theta dt d\phi-4n^{2}N^{2}\cos^{2}\theta d\phi^{2}
+N^{-2} dr^{2}+\Sigma(d\theta^{2}+\sin^{2}\theta d\phi^{2}),
\label{taubmetric2}
\eeq
where
\beq
N^{2}=\frac{\Delta}{\Sigma}=\frac{(r-r_{+})(r-r_{-})[r^{2}+(r_{+}+r_{-})r-c]}{l^{2}(r^{2}+n^{2})}.
\label{lapse1}
\eeq
Now, we describe how to construct the GEMS structure for the Taub-NUT-AdS spacetime, 
because the GEMS has a highly nontrivial form. In order to generate the $dt d\phi$-term 
in (\ref{taubmetric2}), we take an ansatz for $(z^{0},...,z^{3})$ in the GEMS for the Taub-NUT spacetime:
\bea
z^{0}&=&\kappa^{-1}N\cos\frac{\theta}{2}\sinh\kappa (t+2n\phi),\nn\\
z^{1}&=&\kappa^{-1}N\cos\frac{\theta}{2}\cosh\kappa (t+2n\phi),\nn\\
z^{2}&=&\kappa^{-1}N\sin\frac{\theta}{2}\sinh\kappa (t-2n\phi),\nn\\
z^{3}&=&\kappa^{-1}N\sin\frac{\theta}{2}\cosh\kappa (t-2n\phi),\label{0123}
\eea
from which we find
\beq
-(dz^{0})^{2}+(dz^{1})^{2}-(dz^{2})^{2}+(dz^{3})^{2}
=ds^{2}-\left[N^{-2}-\kappa^{-2}\left(\frac{dN}{dr}\right)^{2}\right]dr^{2}
-\left(\Sigma-\frac{1}{4}\kappa^{-2}N^{2}\right)d\theta^{2}-\tilde{\Sigma}\sin^{2}\theta d\phi^{2}.
\label{d0123}
\eeq 
Here $\tilde{\Sigma}$ is defined as 
\beq
\tilde{\Sigma}=\Sigma+4n^{2}N^{2}.
\label{tildes}
\eeq
With another ansatz
\bea
z^{4}&=&\tilde{\Sigma}^{1/2}\sin\theta\cos\phi,\nn\\
z^{5}&=&\tilde{\Sigma}^{1/2}\sin\theta\sin\phi,\nn\\
z^{6}&=&\tilde{\Sigma}^{1/2}\cos\theta,\nn\\
z^{7}&=&\left(4n^{2}+\frac{1}{4}\kappa^{-2}\right)^{1/2}N\cos\theta,\nn\\
z^{8}&=&\left(4n^{2}+\frac{1}{4}\kappa^{-2}\right)^{1/2}N\sin\theta,
\label{45678}
\eea
we produce
\beq
(dz^{4})^{2}+(dz^{5})^{2}+(dz^{6})^{2}-(dz^{7})^{2}-(dz^{8})^{2}
=\left(\frac{d\tilde{\Sigma}^{1/2}}{dr}\right)^{2}dr^{2}
+\tilde{\Sigma}(d\theta^{2}+\sin^{2}\theta d\phi^{2})
-\left(4n^{2}+\frac{1}{4}\kappa^{-2}\right)\left[\left(\frac{dN}{dr}\right)^{2}dr^{2}
+N^{2}d\theta^{2}\right].
\label{d45678}
\eeq

By adding up the results in (\ref{d0123}) and (\ref{d45678}), we obtain
\bea
ds^{2}&=&-(dz^{0})^{2}+(dz^{1})^{2}-(dz^{2})^{2}+(dz^{3})^{2}
+(dz^{4})^{2}+(dz^{5})^{2}+(dz^{6})^{2}-(dz^{7})^{2}-(dz^{8})^{2}\nn\\
&&+\left[N^{-2}-\kappa^{-2}_{H}\left(\frac{dN}{dr}\right)^{2}-\left(\frac{d\tilde{\Sigma}^{1/2}}{dr}\right)^{2}\right]dr^{2},
\label{dsf}
\eea
where we have exploited the following ansatz for the relation between 
$\kappa$ and $\kappa_{H}$ in (\ref{kappah})
\beq
\kappa^{-2}=\frac{4}{3}\kappa^{-2}_{H}+\frac{16}{3}n^{2}.
\label{kk}
\eeq

After tedious algebra for removing the coordinate singularity at $r=r_{+}$, we obtain
\bwide
\beq
N^{-2}-\kappa^{-2}_{H}\left(\frac{dN}{dr}\right)^{2}
=\frac{l^{2}A(r)B(r)}{(r-r_{-})(r^{2}+n^{2})^{3}[r^{2}+(r_{+}+r_{-})r-c](r_{+}-r_{-})^{2}(2r_{+}^{2}+r_{+}r_{-}-c)^{2}},
\label{nk}
\eeq
where
\bea
A(r)&=&(r^{2}+n^{2})^{2}(r_{+}-r_{-})(2r_{+}^{2}+r_{+}r_{-}-c)+(r_{+}^{2}+n^{2})\{2r^{5}+4n^{2}r^{3}-(r_{+}+r_{-})(r_{+}r_{-}+c)r^{2}\nn\\
&&+[2n^{2}(r_{+}r_{-}-(r_{+}+r_{-})^{2}-c)+2r_{+}r_{-}c]r+n^{2}(r_{+}+r_{-})(r_{+}r_{-}+c)\},\nn\\
B(r)&=&-2(r_{+}^{2}+n^{2})r^{4}-[r_{+}r_{-}(r_{+}+r_{-})+(r_{+}-r_{-})c+2n^{2}r_{+}]r^{3}
-[r_{+}^{2}r_{-}(r_{+}+r_{-})+r_{+}(r_{+}-r_{-})c+4n^{4}\nn\\
&&+6n^{2}r_{+}^{2}]r^{2}+\{2r_{+}^{2}r_{-}c-4n^{4}r_{+}-n^{2}[2r_{+}^{3}+r_{+}r_{-}(r_{+}+r_{-})+(r_{+}-3r_{-})c]\}r-2n^{4}(r_{+}^{2}-r_{-}^{2}-r_{+}r_{-}-c)\nn\\
&&+n^{2}r_{+}(r_{+}+r_{-})(r_{+}r_{-}+c).\nn
\eea
Here we emphasize that the ansatz in (\ref{kk}) plays a crucial role in removing the coordinate 
singularity at $r=r_{+}$ in the process to evaluate (\ref{nk}). Next, we obtain 
\beq
\left(\frac{d\tilde{\Sigma}^{1/2}}{dr}\right)^{2}=\frac{C(r)^{2}}{l^{2}(r^{2}+n^{2})^{3}\{l^{2}(r^{2}+n^{2})^{2}
+4n^{2}(r-r_{+})(r-r_{-})[r^{2}+(r_{+}+r_{-})r-c]\}},
\label{tsigma}
\eeq
where
\bea
C(r)&=&l^{2}r(r^{2}+n^{2})^{2}+2n^{2}(r^{2}+n^{2})\{4r^{3}+2[r_{+}r_{-}-(r_{+}+r_{-})^{2}-c]r+(r_{+}+r_{-})(r_{+}r_{-}+c)\}\nn\\
&&-4n^{2}r(r-r_{+})(r-r_{-})[r^{2}+(r_{+}+r_{-})r-c].
\eea
\ewide
Now, we redefine the last piece in (\ref{dsf}) as 
\beq
\left[N^{-2}-\kappa^{-2}_{H}\left(\frac{N}{dr}\right)^{2}-\left(\frac{d\tilde{\Sigma}^{1/2}}{dr}\right)^{2}\right]dr^{2}
=(dz^{9})^{2}-(dz^{10})^{2}.
\eeq
In order to split into positive and negative parts, we reshuffle terms in the numerator in (\ref{nk}) 
to arrive at the (6+5)-dimensional GEMS structure
\beq
ds^{2}=-(dz^{0})^{2}+(dz^{1})^{2}-(dz^{2})^{2}+(dz^{3})^{2}+(dz^{4})^{2}
+(dz^{5})^{2}+(dz^{6})^{2}-(dz^{7})^{2}-(dz^{8})^{2}
+(dz^{9})^{2}-(dz^{10})^{2},\nn\\
\label{gemstnads}
\eeq
where $(z^{0},...,z^{8})$ are given by (\ref{0123}) and (\ref{45678}) together with (\ref{kk}). 
The $(z^{9},z^{10})$ are found to become
\bwide
\bea
z^{9}&=&\int dr~\left[\frac{l^{2}\{[c_{1}(r)-cc_{2}(r)][c_{5}(r)-cc_{6}(r)]+[c_{3}(r)-cc_{4}(r)]
[c_{7}(r)-cc_{8}(r)]\}}{(r-r_{-})(r^{2}+n^{2})^{3}[r^{2}+(r_{+}+r_{-})r-c](r_{+}-r_{-})^{2}(2r_{+}^{2}+r_{+}r_{-}-c)^{2}}\right]^{1/2},\nn\\
z^{10}&=&\int dr~\left[\frac{l^{2}\{[(c_{1}(r)-cc_{2}(r)][c_{7}(r)-cc_{8}(r)]+[c_{3}(r)-cc_{4}(r)]
[c_{5}(r)-cc_{6}(r)]\}}{(r-r_{-})(r^{2}+n^{2})^{3}[r^{2}+(r_{+}+r_{-})r-c](r_{+}-r_{-})^{2}(2r_{+}^{2}+r_{+}r_{-}-c)^{2}}
+\left(\frac{d\tilde{\Sigma}^{1/2}}{dr}\right)^{2}\right]^{1/2},
\label{9101}
\eea
\ewide
where
\bea
c_{1}(r)&=&2(r^{2}+n^{2})^{2}(r_{+}-r_{-})r_{+}^{2}+(r_{+}^{2}+n^{2})[2r^{5}+4n^{2}r^{3}-(r_{+}+r_{-})r_{+}r_{-}r^{2}],\nn\\
c_{2}(r)&=&(r^{2}+n^{2})^{2}(r_{+}-r_{-})+(r_{+}^{2}+n^{2})[(r_{+}+r_{-})r^{2}+2(n^{2}-r_{+}r_{-})r],\nn\\
c_{3}(r)&=&n^{2}(r_{+}^{2}+n^{2})\{2[(r_{+}+r_{-})^{2}-r_{+}r_{-}]r-(r_{+}+r_{-})r_{+}r_{-}\}-(r^{2}+n^{2})^{2}(r_{+}-r_{-})r_{+}r_{-},\nn\\
c_{4}(r)&=&n^{2}(r_{+}^{2}+n^{2})(r_{+}+r_{-}),\nn\\
c_{5}(r)&=&-r_{+}r_{-}(r_{+}+r_{-})r^{3}-r_{+}^{2}r_{-}(r_{+}+r_{-})r^{2}-n^{2}r_{+}r_{-}(r_{+}+r_{-})r+2n^{4}r_{-}^{2},\nn\\
c_{6}(r)&=&(r_{+}-r_{-})r^{3}+r_{+}(r_{+}-r_{-})r^{2}+[-2r_{+}^{2}r_{-}+n^{2}(r_{+}-3r_{-})]r,\nn\\
c_{7}(r)&=&2(r_{+}^{2}+n^{2})r^{4}+2n^{2}r_{+}r^{3}+2n^{2}(3r_{+}^{2}+2n^{2})r^{2}+2n^{2}r_{+}(r_{+}^{2}+2n^{2})r
+2n^{4}r_{+}(r_{+}-r_{-})-n^{2}r_{+}^{2}r_{-}(r_{+}+r_{-}),\nn\\
c_{8}(r)&=&n^{2}[2n^{2}+r_{+}(r_{+}+r_{-})],\label{cccc}
\eea
and $\left(d\tilde{\Sigma}^{1/2}/dr\right)^{2}$ is given by (\ref{tsigma}). In our analysis 
we exclude the possibility of $r = r_{-}$, because at that point $(z^{9}, z^{10})$ in (\ref{9101}) is singular.
We have imposed a restriction that $r_{-}<0$ and $r_{+}+r_{-}>0$. Moreover, in the vanishing $l^{-2}$ limit, as in (\ref{rpmads}) $r_{-}$ becomes negative  and it satisfies $r_{+}+r_{-}>0$. The restriction implies that the roots $r_{+}$ and $r_{-}$ in the case of the Taub-NUT-AdS are deformed continuously from the Taub-NUT to maintain the condition $r_{-}<0$ and $r_{+}+r_{-}>0$. From now on, 
we will keep the above restriction and we will use the same notations for $r_{\pm}$ as those for the case of the Taub-NUT-AdS, even though the values of $r_{\pm}$ are different depending on the four-metrics of interest such as the Taub-NUT metric. 


Next, for the case of $0< r < r_{+}$ with $r_{-}<0$ and $r_{+}+r_{-}>0$, the four-metric (\ref{taubmetric2}) is modified as 
\beq
ds^{2}=\bar{N}^{2}dt^{2}+4n\bar{N}^{2}\cos\theta dt d\phi+4n^{2}\bar{N}^{2}\cos^{2}\theta d\phi^{2}
-\bar{N}^{-2} dr^{2}+\Sigma(d\theta^{2}+\sin^{2}\theta d\phi^{2}),
\label{taubmetric3}
\eeq
where
\beq
\bar{N}^{2}=\frac{(r_{+}-r)(r-r_{-})[r^{2}+(r_{+}+r_{-})r-c]}{l^{2}(r^{2}+n^{2})}.
\eeq
We thus readily find (8+3) GEMS structure 
\beq
ds^{2}=+(dz^{0})^{2}-(dz^{1})^{2}+(dz^{2})^{2}-(dz^{3})^{2}+(dz^{4})^{2}+(dz^{5})^{2}+(dz^{6})^{2}+(dz^{7})^{2}+(dz^{8})^{2}
+(dz^{9})^{2}-(dz^{10})^{2},
\label{gemstnadsinside}
\eeq
with coordinate transformations:
\bea
z^{0}&=&\kappa^{-1}\bar{N}\cos\frac{\theta}{2}\sinh\kappa (t+2n\phi),\nn\\
z^{1}&=&\kappa^{-1}\bar{N}\cos\frac{\theta}{2}\cosh\kappa (t+2n\phi),\nn\\
z^{2}&=&\kappa^{-1}\bar{N}\sin\frac{\theta}{2}\sinh\kappa (t-2n\phi),\nn\\
z^{3}&=&\kappa^{-1}\bar{N}\sin\frac{\theta}{2}\cosh\kappa (t-2n\phi),\nn\\
z^{4}&=&(\Sigma-4n^{2}\bar{N}^{2})^{1/2}\sin\theta\cos\phi,\nn\\
z^{5}&=&(\Sigma-4n^{2}\bar{N}^{2})^{1/2}\sin\theta\sin\phi,\nn\\
z^{6}&=&(\Sigma-4n^{2}\bar{N}^{2})^{1/2}\cos\theta,\nn\\
z^{7}&=&\left(4n^{2}+\frac{1}{4}\kappa^{-2}\right)^{1/2}\bar{N}\cos\theta,\nn\\
z^{8}&=&\left(4n^{2}+\frac{1}{4}\kappa^{-2}\right)^{1/2}\bar{N}\sin\theta,
\label{010inside}
\eea
and $(z^{9},z^{10})$ is given by (\ref{9101}). 

\section{Parameter reductions of Taub-NUT-AdS GEMS}
\setcounter{equation}{0}
\renewcommand{\theequation}{\arabic{section}.\arabic{equation}}

\subsection{Taub-NUT limit}

Now, we investigate a Taub-NUT spacetime limit by exploiting the parameter reduction process 
associated with vanishing $l^{-2}$. Using the parameter reduction in (\ref{delta}) and (\ref{sigma}), we find for $r\geq r_{+}$
\beq
\Delta_{TN}=r^{2}-2Mr-n^{2},~~~\Sigma_{TN}=\Sigma=r^{2}+n^{2},
\label{deltatn}
\eeq
to yield the Taub-NUT spacetime metric
\beq
ds_{TN}^{2}=-N_{TN}^{2}dt^{2}-4nN_{TN}^{2}\cos\theta dt d\phi-4n^{2}N_{TN}^{2}\cos^{2}\theta d\phi^{2}
+N_{TN}^{-2} dr^{2}+\Sigma_{TN}(d\theta^{2}+\sin^{2}\theta d\phi^{2}).
\label{taubmetric32}
\eeq
Here we find
\beq
N_{TN}^{2}=\frac{(r-r_{+})(r-r_{-})}{(r^{2}+n^{2})},
\label{ntn}
\eeq
where $r_{\pm}$ are given by
\beq
r_{\pm}=M\pm (M^{2}+n^{2})^{1/2}.
\label{rpmads}
\eeq
Here one notes that $r_{+}$ is an event horizon and $r_{-}$ is negative. Moreover, we obtain in the vanishing $l^{-2}$,
\bea
\tilde{\Sigma}_{TN}&=&\Sigma_{TN}+4n^{2}N_{TN}^{2},\label{stn}\\
(\kappa^{TN})^{-2}&=&
\frac{16}{3}\left[\left(\frac{(r_{+})^{2}+n^{2}}{r_{+}-r_{-}}\right)^{2}+n^{2}\right].
\label{ktn}
\eea

We proceed to exploit the parameter reduction and we find modified coordinate transformations 
associated with the (6+5) GEMS structure in (\ref{gemstnads}). The coordinate transformations 
$(z^{0},...,z^{8})$ are obtained 
by replacing the $N$, $\tilde{\Sigma}$ and $\kappa$ with $N_{TN}$ in (\ref{ntn}), $\tilde{\Sigma}_{TN}$ 
in (\ref{stn}) and $\kappa^{TN}$ in (\ref{ktn}), respectively. Moreover, we find the 
coordinate transformations  for $(z^{9},z^{10})$ as follows
\bea
z^{9}&=&\int dr~\left[\frac{c_{2}(r)c_{6}(r)+c_{4}(r)c_{8}(r)}
{(r-r_{-})(r^{2}+n^{2})^{3}(r_{+}-r_{-})^{2}}\right]^{1/2},\nn\\
z^{10}&=&\int dr~\left[\frac{c_{2}(r)c_{8}(r)+c_{4}(r)c_{6}(r)}
{(r-r_{-})(r^{2}+n^{2})^{3}(r_{+}-r_{-})^{2}}+\left(\frac{d\tilde{\Sigma}_{TN}^{1/2}}{dr}\right)^{2}\right]^{1/2},
\label{9102}
\eea
where $c_{i}(r)~(i=2,4,6,8)$ can be read off from (\ref{cccc}) and $\left(d\tilde{\Sigma}_{TN}^{1/2}/dr\right)^{2}$ is given by 
\bwide
\beq
\left(\frac{d\tilde{\Sigma}_{TN}^{1/2}}{dr}\right)^{2}=\frac{\{r(r^{2}+n^{2})^{2}+2n^{2}(r^{2}+n^{2})[2r-(r_{+}+r_{-})]
-4n^{2}r(r-r_{+})(r-r_{-})\}^{2}}{(r^{2}+n^{2})^{3}\{(r^{2}+n^{2})^{2}
+4n^{2}(r-r_{+})(r-r_{-})\}}.
\label{tsigma2}
\eeq
\ewide
The above GEMS results can be also evaluated~\footnote{In Ref.~\cite{hong10}, there are typographical errors in expressions 
for $(z^{9},z^{10})$ of the Taub-NUT GEMS.} by making a start from 
the Taub-NUT metric (\ref{taubmetric32}) without the parameter reduction process.

Next, for the case of $0< r < r_{+}$, we find modified coordinate transformations 
associated with the (8+3) GEMS structure in (\ref{gemstnadsinside}). The coordinate transformations 
$(z^{0},...,z^{8})$ are obtained 
by replacing the $N_{TN}^{2}$ in (\ref{ntn}) and $\Sigma_{TN}+4n^{2}N_{TN}^{2}$ with $\bar{N}^{2}_{TN}$ 
and $\Sigma_{TN}-4n^{2}\bar{N}_{TN}^{2}$, respectively, where 
\beq
\bar{N}^{2}_{TN}=\frac{(r_{+}-r)(r-r_{-})}{(r^{2}+n^{2})}.
\label{ntninside}
\eeq 
The coordinate transformations  for $(z^{9},z^{10})$ are given by those in (\ref{9102}).

\subsection{Schwarzschild-AdS limit}

In order to investigate Schwarzschild-AdS spacetime limit, we perform the parameter reduction process 
associated with vanishing $n$. Using the parameter reduction in (\ref{delta}) we find 
the Schwarzschild-AdS spacetime metric
\beq
ds_{SA}^{2}=-N_{SA}^{2}dt^{2}+N_{SA}^{-2} dr^{2}+r^{2}(d\theta^{2}+\sin^{2}\theta d\phi^{2}),
\label{sa2}
\eeq
where
\beq
N_{SA}^{2}=1-\frac{2M}{r}+\frac{r^{2}}{l^{2}}.
\label{nsa}
\eeq
Here three roots of $N_{SA}^{2}=0$ are given by one positive real root and two imaginary ones: 
\bea
r&=&r_{H},\nn\\
r&=&r_{1,2}=\frac{1}{2}\left[-r_{H}\pm i (3r_{H}^{2}+4l^{2})^{1/2}\right].\nn
\eea
An algorithm for roots of a cubic equation is summarized in Refs.~\cite{weisstein03,hong13}.

Next, we obtain in the vanishing $n$ limit of (\ref{sigma}), 
(\ref{kappah}) and (\ref{tildes}),
\bea
\tilde{\Sigma}_{SA}&=&\Sigma_{SA}=r^{2},\label{ssa}\\
\kappa^{SA}&=&\kappa_{H}^{SA}=\frac{3r_{H}^{2}+l^{2}}{2l^{2}r_{H}}.
\label{ksa}
\eea
Because the Schwarzschild-AdS metric does not contain $dt d\phi$-term, we do not need four 
coordinate transformations as in (\ref{0123}). Instead, we take an ansatz for $(z^{0},z^{1})$ for $r\geq r_{H}$:
\bea
z^{0}&=&(\kappa_{H}^{SA})^{-1}N_{SA}\sinh\kappa_{H}^{SA}t,\nn\\
z^{1}&=&(\kappa_{H}^{SA})^{-1}N_{SA}\cosh\kappa_{H}^{SA}t.
\label{01sa}
\eea
Similarly, in this limit, $(z^{4},...,z^{8})$ in (\ref{45678}) merge into $(z^{2},z^{3},z^{4})$ as follows
\bea
z^{2}&=&r\sin\theta\cos\phi,\nn\\
z^{3}&=&r\sin\theta\sin\phi,\nn\\
z^{4}&=&r\cos\theta,
\label{234sa}
\eea
where we have used the identity (\ref{ssa}).

By exploiting the results in (\ref{01sa}) and (\ref{234sa}), we obtain
\beq
ds_{SA}^{2}=-(dz^{0})^{2}+(dz^{1})^{2}+(dz^{2})^{2}+(dz^{3})^{2}+(dz^{4})^{2}
+\left[N^{-2}-(\kappa_{H}^{SA})^{-2}\left(\frac{dN_{SA}}{dr}\right)^{2}-1\right]dr^{2}.
\label{dsfsa}
\eeq
Defining the last term in (\ref{dsfsa}) as 
\beq
\left[N^{-2}-(\kappa_{H}^{SA})^{-2}\left(\frac{dN_{SA}}{dr}\right)^{2}-1\right]dr^{2}=(dz^{5})^{2}-(dz^{6})^{2}
\eeq
we arrive at~\cite{kim00,deser97,deser98}
\bea
z^{5}&=&\int dr~\left[\frac{l^{2}(r_{H}^{2}+l^{2})^{2}r_{H}(r^{2}+r_{H}r+r_{H}^{2})}{(3r_{H}^{2}+l^{2})^{2}r^{3}(r^{2}+r_{H}r+r_{H}^{2}+l^{2})}\right]^{1/2},\nn\\
z^{6}&=&\int dr~\left[\frac{(9r_{H}^{4}+10r_{H}^{2}l^{2}+l^{4})(r^{2}+r_{H}r+r_{H}^{2})}{(3r_{H}^{2}+l^{2})^{2}(r^{2}+r_{H}r+r_{H}^{2}+l^{2})}\right]^{1/2}.\nn\\
\label{56sa}
\eea
At this stage, the parameter reduction scheme used in the case of Taub-NUT limit is not applicable to the 
Schwarzschild-AdS case, because the Taub-NUT possesses at least two real roots $(r_{+},r_{-})$ while the 
Schwarzschild-AdS has one real and two imaginary roots $(r_{+},r_{1},r_{2})$. Finally we find 
the (5+2) GEMS structure:
\beq
ds_{SA}^{2}=-(dz^{0})^{2}+(dz^{1})^{2}+(dz^{2})^{2}+(dz^{3})^{2}+(dz^{4})^{2} +(dz^{5})^{2}-(dz^{6})^{2}.
\label{gemssa}
\eeq

Finally, for the case of $r<r_{H}$, the four-metric (\ref{sa2}) is modified as 
\beq
ds_{SA}^{2}=\bar{N}_{SA}^{2}dt^{2}-\bar{N}_{SA}^{-2} dr^{2}+r^{2}(d\theta^{2}+\sin^{2}\theta d\phi^{2}),
\label{sainside}
\eeq
where
\beq
\bar{N}_{SA}^{2}=-1+\frac{2M}{r}-\frac{r^{2}}{l^{2}}.
\label{nsa2}
\eeq
We thus readily find (5+2) GEMS structure 
\beq
ds_{SA}^{2}=+(dz^{0})^{2}-(dz^{1})^{2}+(dz^{2})^{2}+(dz^{3})^{2}+(dz^{4})^{2} +(dz^{5})^{2}-(dz^{6})^{2}.
\label{gemssainside}
\eeq
with coordinate transformations:
\bea
z^{0}&=&(\kappa_{H}^{SA})^{-1}\bar{N}_{SA}\sinh\kappa_{H}^{SA}t,\nn\\
z^{1}&=&(\kappa_{H}^{SA})^{-1}\bar{N}_{SA}\cosh\kappa_{H}^{SA}t,\nn
\eea
and $(z^{2},...,z^{6})$ are given by those in (\ref{234sa}) and (\ref{56sa}). 

\subsection{Schwarzschild limit}

In the limit of vanishing $n$ and $l^{-2}$, we find the lapse function for the Schwarzschild black hole:
\beq
N_{S}^{2}=1-\frac{2M}{r}.
\label{ns}
\eeq
Applying the parameter reduction scheme to the Schwarzschild-AdS in vanishing $l^{-2}$ limit, one finds
that the coordinate transformations for $(z^{5},z^{6})$ merge into $z^{5}$ for $r\geq r_{H}$~\cite{kim00,deser97,deser98}
\beq
z^{5}=\int dr~\left[\frac{r_{H}(r^{2}+r_{H}r+r_{H}^{2})}{r^{3}}\right]^{1/2},
\label{5s}
\eeq
with $r_{H}=2M$. The other coordinate transformations can be readily read off from (\ref{01sa}) and (\ref{234sa}),
to yield (5+1)-dimensional GEMS structure.

Finally, for the case of $r<r_{H}$, we find modified coordinate transformations 
associated with the (5+1) GEMS structure. The coordinate transformations 
$(z^{0},z^{1})$ are obtained 
by replacing the $N_{S}^{2}$ in (\ref{ns}) with $\bar{N}^{2}_{S}$ where 
$$
\bar{N}^{2}_{S}=-1+\frac{2M}{r}.
$$
The coordinate transformation for $z^{5}$ is given by that in (\ref{5s}) 
and $(z^{2},z^{3},z^{4})$ are given by those in (\ref{234sa}).

\section{Conclusions}
\setcounter{equation}{0}
\renewcommand{\theequation}{\arabic{section}.\arabic{equation}}

We have constructed a global flat embedding of a Taub-NUT-AdS spacetime to yield (6+5) GEMS. 
In the vanishing cosmological constant limit, we have formulated a Taub-NUT GEMS using 
the parameter reduction algorithm. In the vanishing monopole strength limit of the Taub-NUT-AdS, the parameter reduction has been shown not to be successfully applicable to the Schwarzschild-AdS, because there exist 
imaginary roots of Schwarzschild-AdS lapse function associated with event horizon. It has been also 
shown in the vanishing cosmological constant limit that the Schwarzschild GEMS is readily 
formulated by applying the parameter reduction to the Schwarzschild-AdS. 
Finally, for the Taub-NUT-AdS, Taub-NUT, Schwarzschild-AdS and 
Schwarzschild manifolds, we have constructed the GEMS for the patches inside the event horizons of these manifolds. 
It will be interesting to extend the results of the Taub-NUT-AdS GEMS 
to the Taub-Bolt-AdS metric~\cite{hong14}.


\begin{thebibliography}{99}
\bibitem{taub51} A. Taub, Ann. Math. 53, 472 (1951).
\bibitem{nut63} E.T. Newman, L. Tamburino, and T. Unti, J. Math. Phys. 4, 915 (1963).
\bibitem{witten09} E. Witten, JHEP 0906, 067 (2009). 
\bibitem{jezierski07} J. Jezierski and M. Lukasik, Class. Quant. Grav. 24, 1331 (2007). 
\bibitem{cherkis09} S.A. Cherkis, Commun. Math. Phys. 290, 719 (2009). 
\bibitem{kunz10} V. Kagramanova, J. Kunz, E. Hackmann and C. Lammerzahl, Phys. Rev. D 81, 124044 (2010). 
\bibitem{bena11} I. Bena, S. Giusto and C. Ruef, JHEP 1106, 140 (2011). 
\bibitem{beyer12} F. Beyer and J. Hennig, Class. Quant. Grav. 29, 245017 (2012).
\bibitem{moroianu11} A. Moroianu and S. Moroianu, Commun. Math. Phys. 305, 641 (2011).
\bibitem{jante14} R. Jante and B.J. Schroers, JHEP 1401, 114 (2014). 
\bibitem{bardoux14} Y. Bardoux, M.M. Caldarelli and C. Charmousis, JHEP 1405, 039 (2014). 
\bibitem{hawking77} S.W. Hawking, Phys. Lett. A 60, 81 (1977).
\bibitem{manton85} N.S. Manton, Phys. Lett. B 110, 54 (1985).
\bibitem{atiyah85} M.F. Atiyah and N. Hitchin, Phys. Lett. A 107, 21 (1985).
\bibitem{awad02} A. Awad and A. Chamblin, Class. Quant. Grav. 19, 2051 (2002).
\bibitem{bini03} D. Bini, C. Cherubini, M. de Mattia and R.T. Jantzen, Gen. Rel. Grav. 35, 2249 (2003); D. Bini, C. Cherubini, R.T.
Jantzen and B. Mashhoon, Class. Quantum Grav. 20, 457 (2003); D. Bini, C. Cherubini, R.T. Jantzen and B. Mashhoon,
Phys. Rev. D 67, 084013 (2003).
\bibitem{jezierski072} J. Jezierski and M. Lukasik, Class. Quant. Grav. 24, 1331 (2007). 
\bibitem{kasner21} E. Kasner, Am. J. Math. 43, 130 (1921).
\bibitem{fronsdal59} C. Fronsdal, Phys. Rev. 116, 778 (1959).
\bibitem{friedman65} A. Friedman, Rev. Mod. Phys. 37, 201 (1965).
\bibitem{goenner80} H. F. Goenner, General Relativity and Gravitation, edited by A. Held
(Plenum, New York, 1980).
\bibitem{rosen65} J. Rosen, Rev. Mod. Phys. 37, 204 (1965).
\bibitem{narnhofer96} H. Narnhofer, I. Peter, and W. Thirring, Int. J. Mod. Phys. B 10, 1507 (1996).
\bibitem{chamblin97} A. Chamblin and G. W. Gibbons, Phys. Rev. D 55, 2177 (1997).
\bibitem{deser97} S. Deser and O. Levin, Class. Quant. Grav. 14, L163 (1997). 
\bibitem{deser98} S. Deser and O. Levin, Class. Quant. Grav. 15, L85 (1998). 
\bibitem{kim00} Y.W. Kim, Y.J. Park and K.S. Soh, Phys. Rev. D 62, 104020 (2000). 
\bibitem{hong00} S.T. Hong, Y.W. Kim and Y.J. Park, Phys. Rev. D 62, 024024 (2000).
\bibitem{hong002} S.T. Hong, W.T. Kim, Y.W. Kim and Y.J. Park, Phys. Rev. D 62, 064021 (2000).
\bibitem{hong01} S.T. Hong, W.T. Kim, J.J. Oh and Y.J. Park, Phys. Rev. D 63, 127502 (2001).
\bibitem{hong04} S.T. Hong, Gen. Rel. Grav. 36, 1919 (2004).
\bibitem{hong10} S.T. Hong and S.W. Kim, J. Korean Phys. Soc. 56, 1633 (2010). 
\bibitem{banerjee10} R. Banerjee and B.R. Majhi, Phys. Lett. B 690, 83 (2010).
\bibitem{hong12} S.T. Hong, arXiv:1209.4563.
\bibitem{hong132} S.T. Hong, J. Korean Phys. Soc. 64, 1928 (2014).
\bibitem{gems08} E. J. Brynjolfsson and L. Thorlacius, JHEP 0809, 066 (2008).
\bibitem{gems14} Y. W. Kim, J. Choi and Y. J. Park, Phys. Rev. D 89, 044004 (2014).

\bibitem{hawking75} S.W. Hawking, Comm. Math. Phys. 42, 199 (1975).
\bibitem{unruh76} W.G. Unruh, Phys. Rev. D 14, 870 (1976).
\bibitem{chen04} H.Z. Chen, Y. Tian, Y.H. Gao and X.C. Song, JHEP 0410, 011 (2004).
\bibitem{santos04} N.L. Santos, O.J.C. Dias and J.P.S. Lemos, Phys. Rev. D 70, 124033 (2004).
\bibitem{tian05} Y. Tian, JHEP 0506, 045 (2005).
\bibitem{radu06} E. Radu, Int. J. Mod. Phys. A 21, 4355 (2006).
\bibitem{langlois06} P. Langlois, Annals Phys. 321, 2027 (2006).
\bibitem{russo08} J.G. Russo and P.K. Townsend, Class. Quant. Grav. 25, 175017 (2008).
\bibitem{bekenstein73} J.D. Bekenstein, Phys. Rev. D 7, 2333 (1973).
\bibitem{btz} M. Ba\~nados, C. Teitelboim, and J. Zanelli, Phys. Rev. Lett. 69, 1849 (1992); 
M. Ba\~nados, M. Henneaux, C. Teitelboim, and J. Zanelli, Phys. Rev. D 48, 1506 (1993).
\bibitem{carlip95} S. Carlip, Class. Quantum Grav. 12, 2853 (1995).
\bibitem{cangemi93} D. Cangemi, M. Leblanc, and R.B. Mann, Phys. Rev. D 48, 3606 (1993). 
\bibitem{sch16} K. Schwarzschild, Sitzber. Deut. Akad. Wiss. Berlin, KI. Math. Phys. Tech., pp. 189-196 (1916).
\bibitem{rn18} H. Reissner, Ann. Phys. (Leipzig) 50, 106 (1916); 
G. Nordstr\"om, Proc. K. Ned. Akad. Wet. 20, 1238 (1918).
\bibitem{peca18} C.S. Peca and J.P.S. Lemos, Phys. Rev. D 59, 124007 (1999); 
P. Mitra, Phys. Lett. B 459, 119 (1999); 
S.W. Hawking and H.S. Reall, Phys. Rev. D 61, 024014 (1999); 
B. Wang, E. Abdalla, and R.K. Su, ibid. Phys. Rev. D 62, 047501 (2000).
\bibitem{weisstein03} E.W. Weisstein, CRC Concise Encyclopedia of Mathematics (Chapman \& Hall/CRC, New York, 2003).
\bibitem{hong13} S.T. Hong and Y. Kim, arXiv:1309.2177. 

\bibitem{hong14} S.T. Hong, in preparation.
\end{thebibliography}
\end{document}